\documentclass[
  floats,
  floatfix,
  prd,
  twocolumn,
  superscriptaddress,
  nofootinbib,
  longbibliography,
]{revtex4-2}
\usepackage{amsfonts,amssymb,amsmath}
\usepackage[caption=false]{subfig} %
\usepackage{graphicx, hyperref, tabularx, dcolumn}
\usepackage{multirow, makecell}
\usepackage{algpseudocode}
\makeatletter
\newenvironment{algorithm}{\@float{algorithm}}{\end@float}
\newcounter{algorithm}
\def\fnum@algorithm{Algorithm~\thealgorithm}
\def\fps@algorithm{tbp}
\def\ftype@algorithm{8}
\def\ext@algorithm{loa}
\makeatother

\usepackage{color}
\usepackage[dvipsnames]{xcolor}
\usepackage[normalem]{ulem}
\usepackage{orcidlink}
\usepackage{tensor, comment}
\usepackage{enumitem}
\usepackage{graphicx}
\usepackage{booktabs}

\newcommand{\e}[1]{\times 10^{#1}}

\newcommand{\athenak}{\texttt{AthenaK}}

\definecolor{ykcolor}{rgb}{0.9, 0.25, 0.25}

\graphicspath{{./Figures}}

\newcommand{\caltech}{\affiliation{Theoretical Astrophysics 350-17, California
Institute of Technology, Pasadena, California 91125, USA}}
\newcommand{\burke}{\affiliation{Walter Burke Institute for Theoretical Physics,
California Institute of Technology, Pasadena, California 91125, USA}}

\begin{document}
\raggedbottom
\author{Samuel Solod \orcidlink{0009-0006-2324-8342}}
\email{solodsamuel@gmail.com}
\affiliation{Department of Physics and Astrophysics, DePaul University, Chicago, Illinois 60614, USA}
\caltech

\author{Yoonsoo Kim \orcidlink{0000-0002-4305-6026}} 
\email{yoonsoo@princeton.edu}
\affiliation{Princeton Center for Theoretical Science, Princeton University, Princeton, New Jersey 08544, USA}
\affiliation{Leinweber Forum for Theoretical Physics, Princeton University, Princeton, New Jersey 08544, USA}

\caltech

\author{Elias R. Most \orcidlink{0000-0002-0491-1210}}
\caltech
\burke

\title{A higher-order smoothness indicator as refinement criterion for Newtonian and relativistic MHD simulations}

\date{\today}

\begin{abstract}
  Adaptive mesh refinement is a widely adopted strategy for grid-based numerical
  simulation codes to mitigate the high computational cost of multi-length scale
  problems. A robust and reliable mesh refinement criterion with minimal
  requirement for problem-specific tuning is critical for efficient and accurate
  simulations. Motivated by the oscillation indicators used in the spectral
  methods literature, we develop a new mesh refinement criterion employing a
  5-point stencil for multi-dimensional magnetohydrodynamics simulations,
  essentially extending the conventional 3-point L{\"o}hner scheme. We implement
  the criterion in the astrophysical fluid dynamics code \athenak~and test it
  across a variety of benchmark problems in Newtonian, special relativistic, and
  general relativistic magnetohydrodynamics, demonstrating its robustness and
  versatility.
\end{abstract}

\maketitle

\section{Introduction} \label{sec:intro}

Partial differential equations describing a physical system are often highly
nonlinear and therefore require numerical approaches to solve them. For example,
the evolution of a gaseous medium and magnetized plasma in astrophysics is
frequently modeled with magnetohydrodynamics (MHD) framework, which can be
extended to special relativistic MHD (SRMHD) and general relativistic MHD
(GRMHD) for extreme environments around compact objects. To cope with the highly
complex and nonlinear character of the governing equations, numerical methods
that can accurately and efficiently solve them have been continuously developed
(see, e.g., Ref. \cite{Marti:2015bkp} for a review).

Many applications in computational astrophysics involve a scale separation in
length that may often span orders of magnitude. In global simulations,
attempting to resolve small-scale features by naively increasing the resolution
everywhere can be prohibitively expensive.
The adaptive mesh refinement (AMR) \cite{Berger:1984zza,BERGER198964,Stout:1997}
technique tackles this issue by dynamically controlling the local grid
resolution, substantially reducing the total computational cost while faithfully
resolving small-scale features. AMR has been widely adopted in computational
astrophysics and numerical relativity codes
\cite[e.g.][]{Fryxell:2000zz,Teyssier:2001cp,Schnetter:2003rb,Mignone:2011fd,KEPPENS2012718,Porth:2016rfi,Zhang2019,Stone:2020}
\cite[also see][and references therein]{Dubey:2014mep}.

A key algorithmic element in AMR is the criterion that flags regions in the
computational domain that need greater resolution (refinement) or that can
safely use less resolution (coarsening). This can either be done based on a
natural scale ordering of the system, or because certain parts of the domain
exhibit physical dynamics of interest, or agnostically based on the estimated
numerical error of the solution at a given resolution. A common strategy is to
monitor the local behavior of the numerical solution by means of the truncation
error or other physical admissibility requirements. The curvature indicator by
\citet{1987CMAME..61..323L} has been a popular choice in finite-volume type
codes in computational astrophysics
\cite[e.g.][]{Fryxell:2000zz,Mignone:2011fd,KEPPENS2012718,Zanotti:2015mia,Porth:2016rfi},
later also applied to the spectral method
\cite{Zanotti:2015mia,Renkhoff:2023nfw}. This criterion looks at the second
derivative of a chosen quantity and normalizes with the average first derivative
across the grid cell.

However, the threshold value and the physical quantities used for the
L{\"o}hner-type indicators in the literature remain highly problem-specific
\cite[e.g.][]{Mignone:2011fd,Shin:2012pt,Iapichino:2017kvp,Hillel:2021cir,Schive:2017sdo,Radia:2021smk,Schinasi-Lemberg:2024dko,Stone:2020,Stone_2026},
and are therefore not used in many applications, which resort to a static
collection of refined meshes. While a refinement criterion that depends on a
priori estimates of the magnitude or spatial scale of the physical variables
(e.g. min--max or slope threshold) can be a supplemental strategy, these
estimates might not be readily available before actually performing a simulation
or no longer applicable once the region of interest evolves into a different
regime and moves out from the preset threshold window. This makes their use both
costly due to the overhead in the extra number of simulations, and inefficient
due to the necessity for careful problem-specific fine-tuning of threshold
values.

In this work, we develop and evaluate a new simple refinement criterion targeted
for finite-volume MHD codes, aiming to operate across Newtonian, SRMHD, and
GRMHD simulations. The motivation for this work is mainly twofold. First, we
attempt to develop a simplified AMR criterion that behaves robustly across a
diverse set of benchmark problems without extensive case-by-case tuning. To our
knowledge, such a systematic investigation has received little attention in the
computational astrophysics literature, although the outcome can greatly help
alleviate the need for labor-intensive experimentation by individual code users.
Second, in lockstep with recent advances in higher-order GRMHD codes
\cite{Most:2019kfe,Kiuchi:2025ksk}, we explore the potential of an AMR criterion
that employs a 5-point stencil, wider than the conventional 3-point stencil
methods.

This paper is organized as follows. In Sec.~\ref{sec:criteria}, we describe the
proposed refinement criterion and its example implementation in the
\athenak~code. In Sec.~\ref{sec:results}, we present the test results on the
benchmark problems in Newtonian MHD, SRMHD, and GRMHD. We conclude and list
several points of discussion in Sec.~\ref{sec:discussion}.

\section{Mesh refinement criteria} \label{sec:criteria}

\subsection{Motivation and basic algorithm} \label{sec:basic-algorithm}

\begin{figure*}
\centering
\tikzset{every picture/.style={x=0.061\linewidth, y=0.24\columnwidth},
  note/.style={anchor=north west, align=left, soln, font=\small}}
\captionsetup[subfloat]{captionskip=15pt}

\subfloat[Flagged to refine \label{fig:stencil-refine}]{%
\begin{tikzpicture}[soln/.style={blue!70!black}]
\pgfmathdeclarefunction{u}{1}{\pgfmathparse{0.7 + 0.05*(#1 * (#1+1.2))
  + 0.3*sin(100*(#1-0.))}}
\draw[soln, thick, smooth, domain=-2:3, samples=80, variable=\t]
    plot ({\t}, {u(\t)});
\draw (-2,0) -- (3,0);
\pgfmathsetmacro{\tick}{0.05} %
\foreach \x/\idx in {-2/i-2, -1/i-1, 0/i, 1/i+1, 2/i+2}
{
    \pgfmathsetmacro{\xc}{\x+0.5}
    \pgfmathsetmacro{\uc}{u(\xc)}
    \draw (\xc,0) -- ++(0,-\tick);
    \node[below, inner sep=2pt] at (\xc,-\tick-0.03) {$x_{\idx}$};
    \ifnum\x=0 %
        \filldraw[soln, fill=white, line width=1.3pt] (\xc,\uc) circle (2.2pt);
    \else
        \fill[soln] (\xc,\uc) circle (2pt);
    \fi
    \node[below, inner sep=1pt] (lbl) at (\xc,\uc-0.13) {$u_{\idx}$};
    \draw[dotted, gray] (\xc,0) -- (lbl.south);
}
\node[note] at (-2.0, 1.8)
  {$\displaystyle \left|\frac{\Delta^{(4)} u}{u_i}\right| > 4^{-2}$};
\end{tikzpicture}
}
\hfill
\subfloat[Maintain the grid level \label{fig:stencil-stay}]{%
\begin{tikzpicture}[soln/.style={blue!70!black}]
\pgfmathdeclarefunction{u}{1}{\pgfmathparse{0.7 + 0.35*sin(40*(#1-0.1))}}
\draw[soln, thick, smooth, domain=-2:3, samples=80, variable=\t]
    plot ({\t}, {u(\t)});
\draw (-2,0) -- (3,0);
\pgfmathsetmacro{\tick}{0.05} %
\foreach \x/\idx in {-2/i-2, -1/i-1, 0/i, 1/i+1, 2/i+2}
{
    \pgfmathsetmacro{\xc}{\x+0.5}
    \pgfmathsetmacro{\uc}{u(\xc)}
    \draw (\xc,0) -- ++(0,-\tick);
    \node[below, inner sep=2pt] at (\xc,-\tick-0.03) {$x_{\idx}$};
    \ifnum\x=0 %
        \filldraw[soln, fill=white, line width=1.3pt] (\xc,\uc) circle (2.2pt);
    \else
        \fill[soln] (\xc,\uc) circle (2pt);
    \fi
    \node[below, inner sep=1pt] (lbl) at (\xc,\uc-0.13) {$u_{\idx}$};
    \draw[dotted, gray] (\xc,0) -- (lbl.south);
}
\node[note] at (-2.0, 1.8)
  {$\displaystyle \left|\frac{\Delta^{(4)} u}{u_i}\right| \in [4^{-3}, 4^{-2}]$};
\end{tikzpicture}
}
\hfill
\subfloat[Flagged to coarsen \label{fig:stencil-coarsen}]{%
\begin{tikzpicture}[soln/.style={blue!70!black}]
\pgfmathdeclarefunction{u}{1}{\pgfmathparse{0.7 + 0.5*sin(16*(#1))}}
\draw[soln, thick, smooth, domain=-2:3, samples=80, variable=\t]
    plot ({\t}, {u(\t)});
\draw (-2,0) -- (3,0);
\pgfmathsetmacro{\tick}{0.05} %
\foreach \x/\idx in {-2/i-2, -1/i-1, 0/i, 1/i+1, 2/i+2}
{
    \pgfmathsetmacro{\xc}{\x+0.5}
    \pgfmathsetmacro{\uc}{u(\xc)}
    \draw (\xc,0) -- ++(0,-\tick);
    \node[below, inner sep=2pt] at (\xc,-\tick-0.03) {$x_{\idx}$};
    \ifnum\x=0 %
        \filldraw[soln, fill=white, line width=1.3pt] (\xc,\uc) circle (2.2pt);
    \else
        \fill[soln] (\xc,\uc) circle (2pt);
    \fi
    \node[below, inner sep=1pt] (lbl) at (\xc,\uc-0.13) {$u_{\idx}$};
    \draw[dotted, gray] (\xc,0) -- (lbl.south);
}
\node[note] at (-2.0, 1.8)
  {$\displaystyle \left|\frac{\Delta^{(4)} u}{u_i}\right| < 4^{-3}$};
\end{tikzpicture}
}
\caption{Illustration of the refinement criterion developed in this paper. On
  each grid cell $x_i$, we evaluate the undivided 4th numerical derivative
  $\Delta^{(4)} u \equiv u_{i+2}-4u_{i+1}+6u_i-4u_{i-1}+u_{i-2}$ on a symmetric
  five-point stencil and compare its magnitude against the cell value $|u_i|$.
  Depending on the ratio, the cell is flagged for refinement
  (\ref{fig:stencil-refine}), as neutral (\ref{fig:stencil-stay}), or to coarsen
  (\ref{fig:stencil-coarsen}).}
  \label{fig:criteria}
\end{figure*}
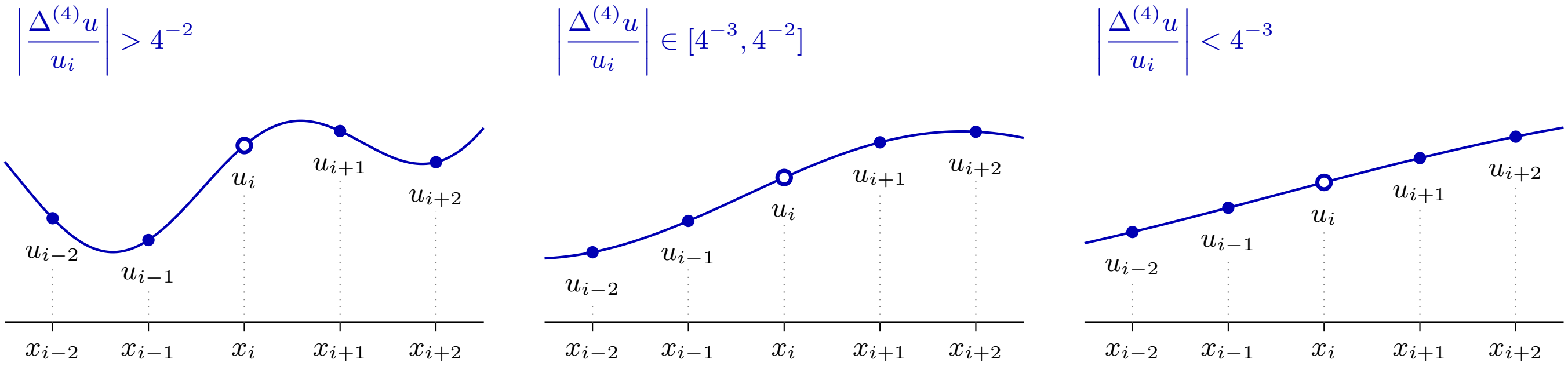

Fig.~\ref{fig:criteria} summarizes the refinement and coarsening criteria that
we propose in this work.
At each grid point $x_i$, we evaluate the local oscillation measure $e_i$
\begin{subequations}
\label{eq:indicator}
\begin{align}
  e_i &=
    \left|\frac{\Delta^{(4)} u}{u_i}\right| \, , \\[1ex]
  \Delta^{(4)} u &\equiv u_{i+2}-4u_{i+1}+6u_i-4u_{i-1}+u_{i-2}  \, ,
\end{align}
\end{subequations}
where $\{u_{i-2}, \cdots, u_{i+2}\}$ are the values of a physical quantity on
the grid stencil $\{x_{i-2}, \cdots, x_{i+2}\}$.
The mesh refinement criteria and thresholds are
\begin{equation} \label{eq:criteria}
  \text{if} \quad
  e_i \,
  \begin{cases}
    \,\, > 4^{-2} & \text{, refine} \\
    \,\, \in [4^{-3}, 4^{-2}]   & \text{, stay} \\
    \,\, < 4^{-3}  & \text{, coarsen}
  \end{cases} \, .
\end{equation}
Note that the threshold for coarsening is set to be lower (tighter) than the
threshold for refinement. Having this gap between the two thresholds prevents
the grid level switching back and forth too frequently at a given location,
which can significantly degrade the accuracy of the solution
\cite{Li_2010,Stone:2020}.\footnote{A similar strategy has been adopted in the
discontinuous Galerkin-finite difference hybrid method
\cite{Deppe:2021bhi,Deppe:2021ada,Kim:2024mau} to reduce the computational cost
and discretization error from the inter-grid projections.} In the following
paragraphs, we outline the motivation and derivation of the criterion
\eqref{eq:indicator} and the thresholds \eqref{eq:criteria}. 

Our criterion borrows its idea from the oscillation indicators used in the
spectral method literature. Spectral methods are prone to nonphysical
high-frequency oscillations (i.e. Gibbs phenomenon) when a steep gradient or
discontinuity---such as shocks or contact discontinuities---is present in the
spectral approximant. A number of different mitigation strategies have been
developed
\cite[e.g.][]{DUMBSER201447,ZANOTTI2015204,VILAR2019245,Deppe:2021ada,RUEDARAMIREZ2022105627,Velasco-Romeri:2023,Adhikari:2025nio},
where many of them invoke a smoothness indicator that monitors the Gibbs
phenomenon.
A common type of the oscillation indicator is monitoring the fall-off rate of
the modal coefficients, or equivalently, the power norm of the solution
contained in the highest-order spectral mode compared to its whole norm
\cite[e.g.][]{Persson,Szilagyi:2014fna,Deppe:2021bhi,Deppe:2023qxa}. The local
behavior of the numerical solution is monitored by means of a dimensionless
number, reducing the need for problem-dependent tuning for the indicator
threshold.\footnote{Interestingly but not so surprisingly, this is also a key
property of the L{\"o}hner-type AMR indicators used in finite volume methods.}
Spectral methods that are supplemented with this modal oscillation detector,
combined with a localized shock-capturing strategy, have been successfully
applied to relativistic MHD
\cite[e.g.][]{Zanotti:2015mia,Deppe:2021bhi,Legred:2023zet,Deppe:2024ckt}
and force-free electrodynamics \cite{Kim:2024mau} simulations.

Specifically, the troubled cell indicator proposed in
\citet{Deppe:2021ada,Deppe:2021bhi} monitors the ratio between the $L_2$ norm in
the highest Legendre polynomial mode and that of the whole polynomial
approximant. A spectral element is marked as troubled if
\begin{equation} \label{eq:tci-criteria}
  \left(\frac{\sum_{i=0}^N \hat{u}_i^2}{\sum_{i=0}^N u_i^2}\right)^{1/2} > N^{-\alpha} \, ,
\end{equation}
with the indicator sensitivity $\alpha=4.0$ \cite{Deppe:2021ada,Deppe:2021bhi}.
Here $u_i$ and $\hat{u}_i$ are the nodal values of the spectral approximant and
its highest mode component at the collocation points $\{x_i\}$. If the left hand
side of Eq.~\eqref{eq:tci-criteria} is smaller than $N^{-(\alpha+1)}$, the
spectral element is deflagged.

We translate \eqref{eq:tci-criteria} into a difference form as follows. Consider
the grid points $\{x_{-N/2}, \cdots, x_0, \cdots x_{N/2}\}$ uniformly spaced
over the interval $[-1, 1]$ for an even number $N$. An $N$-th degree polynomial
approximant to the solution $u(x)$ can be represented in the modal and power
series
\begin{equation} \label{eq:spectral-approximant}
  u(x) = \sum_{j=0}^N c_j P_j(x) = \sum_{k=0}^N a_k x^k \, ,
\end{equation}
where $P_j(x)$ are the Legendre polynomials. The squared norm of the highest
Legendre mode is $\sum_i \hat{u}_i^2 = c_N^2 \sum_i P_N(x_i)^2$, and we
approximate the total squared norm as $\sum_i u_i^2 \approx (N+1) u_0^2$.
Plugging these into \eqref{eq:tci-criteria}, we get
\begin{equation} \label{eq:tci-criteria-approximate}
  |c_N| > N^{-\alpha}
    \sqrt{\frac{N+1}{\sum_i P_N (x_i)^2}}  |u_0| \, .
\end{equation}
Taking the $N$-th derivative of Eq.~\eqref{eq:spectral-approximant} and using
$a_N = c_N (2N)! / (2^N (N!)^2)$, we get
\begin{equation} \label{eq:highest-derivative}
  \frac{d^N u}{dx^N} = \frac{(2N)!}{2^N N!} c_N \, .
\end{equation}
Combining \eqref{eq:tci-criteria-approximate} with \eqref{eq:highest-derivative}
and normalizing with the grid spacing $h\equiv2/(N+1)$,
we get
\begin{equation} \label{eq:tci-criteria-approximate-2}
  \bigg| h^N \frac{d^N u}{dx^N} \bigg|
    > N^{-\alpha}
    \frac{(2N)!}{(N+1)^N N!} 
    \sqrt{\frac{N+1}{\sum_i P_N (x_i)^2}} |u_0| \, .
\end{equation}
Replacing $d^N u / dx^N$ with the discrete finite-difference derivative, the
left hand side of \eqref{eq:tci-criteria-approximate-2} reduces to the undivided
$N$-th order finite-difference formula. For $N=4$, which we adopt for the AMR
indicator devised in this work, $x_i = \{0, \pm \frac{2}{5}, \pm \frac{4}{5}\}$
and Eq.~\eqref{eq:tci-criteria-approximate-2} becomes
\begin{equation}
  \bigg|\frac{u_2 - 4u_1 + 6u_0 - 4u_{-1} + u_{-2}}{u_0}\bigg|
  > 0.045 \approx 4^{-2.24} \, ,
\end{equation}
which motivates our choice of the threshold $4^{-2} = 0.0625$ for refinement and
$4^{-3} = 0.015625$ for coarsening.

\subsection{Adaptation to multi-dimensional MHD}
\label{sec:amr-adaptation-to-mhd}

The oscillation measure $e_i$ can be evaluated for different choices of the
physical variable $\{u_i\}$. The particular choice can differ depending on the
problem of interest; for example, the velocity field $v^i$ for resolving a shear
layer, or magnetic field $B^i$ to increase resolution at current sheets.

We simply use the conserved variables evolved in the MHD solver. They are not
only readily accessible within the code, but also turn out to be robust enough
across a wide variety of test problems, as we show later in
Sec.~\ref{sec:results}. We describe the adaptation of the mesh refinement
criteria \eqref{eq:criteria} to the multi-dimensional MHD solvers as follows.
\begin{itemize}[labelsep=1.5ex, leftmargin=1.5em]
  \item A cell is flagged to refine when \emph{any} monitored field exceeds the
    threshold, and de-refine when \emph{all} of them are below the coarsening
    threshold.
  \item In multi-dimensional simulations, the oscillation measure $e_i$ is
    evaluated along each spatial axis and the largest value is compared against
    the threshold.
  \item For the simulations that involve ambient atmosphere or floored regions,
    introducing a cutoff value can prevent mesh refinements triggered in the
    regions that are dynamically unimportant. Cells with the (rest) mass density
    below a prescribed floor value {$D_{\rm floor}$} are ignored and flagged to
    coarsen by default.
    A few times the atmosphere density used in the MHD solver is sufficient for
    a generic purpose, where higher values can be used to be more selective. We
    present an example in Sec.~\ref{sec:3d-grmhd-test}.
\item For the vector-type conserved variables, we use the magnitude of them as a
    single scalar quantity to be monitored. Since the oscillation measure $e_i$
    is normalized with the cell value $|u_i|$, the zero-crossing behavior of
    them can spuriously trigger the mesh refinement.\\[1ex] For the momentum and
    magnetic field variables, the denominator in the oscillation measure
    \eqref{eq:indicator} is modified to $|u_i| + \epsilon$, where $\epsilon$ is
    a user-defined fudge factor that effectively sets the atmosphere values of
    them.
  \begin{itemize}[leftmargin=1em]
    \item For the momentum variables, a natural first choice is
      $\epsilon_{\rm mom} \sim D_{\rm floor} U$ where $U$ is the characteristic
      scale of the fluid velocity in the code units (e.g. $U\sim
      10^{-2}\text{--}10^{-1}$ in GRMHD black hole accretion simulations). In
      practice, setting $\epsilon$ one or two orders of magnitude higher than
      $D_{\rm floor} U$ is a pragmatic strategy to focus on dynamically
      important shear layers.
    \item For the magnetic fields, we suggest the value at which magnetic
      pressure equals the scale of the gas ram pressure at the AMR floor
      density, i.e. $\epsilon_{\rm B} \sim (D_{\rm floor} U^2)^{1/2}$.
  \end{itemize}
\item In general relativistic MHD, there is an issue regarding the potential
    sign change of the conserved variable in a curved spacetime. See
    Sec.~\ref{sec:3d-grmhd-test} for the related discussions and test results.
\end{itemize}

We implement the refinement criteria above in the astrophysical fluid dynamics
code \athenak~\cite{Stone_2026}. \athenak~adopts a block-based AMR design, and
the computational domain is decomposed into a fixed-size collection of
finite-volume cells called \texttt{MeshBlock}s \cite{Stone:2020}. See the
Refs.~\cite{Stone:2020,Stone_2026} for the technical details of the code.
Algorithm~\ref{indicator-algorithm} describes our implementation. If any cell
within a \texttt{MeshBlock} is flagged to refine, that \texttt{MeshBlock} is
flagged to refine. If all the cells are marked to coarsen, the
\texttt{MeshBlock} is flagged for coarsening.

\begin{algorithm}
  \caption{The AMR refinement criterion implemented in \athenak.}
  \label{indicator-algorithm}
  \vspace{1ex}
  \hrule \vspace{1ex}
\begin{algorithmic}[1]
  \Require Evolved variables $(D, M^j, E, B^j)$, density floor $D_{\rm floor}$,
          division-by-zero guard $(\epsilon_{\rm mom}, \epsilon_{\rm B})$ for
          momentum and magnetic fields.
  \Ensure Refinement flags $f_m\in\{-1,0,1\}$ (coarsen, neutral, refine) for all
    \texttt{MeshBlock}s.
  \Statex
  \Function{$\mathcal{E}_{\boldsymbol{i}}$}{$u, \delta$}
    \Statex \Comment{Oscillation measure of field $u$ at cell $\boldsymbol{i}$}
    \State \Return $\max_{d\in\{\text{spatial axes}\}} |\Delta^{(4)}_d u_{\boldsymbol{i}}| \,/\, (|u_{\boldsymbol{i}}| + \delta)$
  \EndFunction
  \Statex
  \ForAll{\texttt{MeshBlock} $m$}
    \State ${\rm Err}_m \gets -\infty$ \Comment{Initialize to \texttt{-DBL\_MAX}}
      \ForAll{grid cells $\boldsymbol{i}$ in \texttt{MeshBlock} $m$}
          \If{$D_{\boldsymbol{i}} > D_{\mathrm{floor}}$}
            \State $e_{\boldsymbol{i}} \gets \mathcal{E}_{\boldsymbol{i}}(D, 0)$
            \State $e_{\boldsymbol{i}}\gets
                \max[e_{\boldsymbol{i}},\, \mathcal{E}_{\boldsymbol{i}}(\sqrt{(M^j)^2}, \epsilon_{\rm mom})]$
            \State $e_{\boldsymbol{i}}\gets
                \max[e_{\boldsymbol{i}},\, \mathcal{E}_{\boldsymbol{i}}(E, 0)]$
            \State $e_{\boldsymbol{i}}\gets
                \max[e_{\boldsymbol{i}},\, \mathcal{E}_{\boldsymbol{i}}(\sqrt{(B^j)^2}, \epsilon_{\rm B})]$
            \State ${\rm Err}_m \gets\max({\rm Err}_m, e_{\boldsymbol{i}})$
          \EndIf
      \EndFor
      \If{${\rm Err}_m > 4^{-2}$}
        \State $f_m\gets 1$
      \ElsIf{${\rm Err}_m < 4^{-3}$}
        \State $f_m\gets -1$
      \Else
        \State $f_m\gets 0$
      \EndIf
  \EndFor
\vspace{1ex}\hrule
\end{algorithmic}
\end{algorithm}

\section{Test Results}
\label{sec:results}

\begin{table*}
\centering
\renewcommand{\arraystretch}{1.2}
\setlength{\tabcolsep}{8pt}
\caption{List of the benchmark problems and their grid setup.}
\label{tab:amr-setup}
\vspace{1ex}
\begin{tabular}{c|l|ccc|c}
    \hline
      Solver & Name  & Domain size & 
      Base grid & AMR level & Figure \\
    \hline\hline
    \multirowcell{8}{Newtonian\\ MHD}
      & Sod shock tube & $[-0.5, 0.5]$ & $64$ & 5 & \ref{fig:sod} \\
      & Shu--Osher tube & $[-1, 1]$ & $64$ & 5 & \ref{fig:shu-osher} \\
      & Kelvin--Helmholtz instability (KHI) & $[-1, 1]^2$ & $128^2$ & 4 & \ref{fig:kh} \\
      & Rayleigh--Taylor instability & $[-0.2, 0.2]\times[-0.6, 0.6]$ & $32\times64$ & 5 & \ref{fig:rt} \\
      & {Liska--Wendroff implosion} &{$[0, 0.3]^2$} & $64^2$ & 5 & {\ref{fig:lwi_3panel}} \\
    & Magnetized blast wave (MBW) & $[-0.5, 0.5]\times[-0.75, 0.75]$ & $32\times48$ & 5 & \ref{fig:mbw} \\
      & Orszag--Tang vortex & $[-0.5, 0.5]^2$ & $64^2$ & 5 & \ref{fig:ot} \\

    \hline
    \multirow{4}{*}{SRMHD}
      & Cylindrical MBW & $[-6, 6]^2$ & $48^2$ & 5 & \ref{fig:srmhd-blast-2d} \\
      & \multirow{2}{*}{Magnetized KHI} &
        \multirow{2}{*}{$[0, 1]\times [-0.25, 0.25]$} & $64\times 32$ & 5 & \multirow{2}{*}{\ref{fig:srmhd-khi}} \\
      & & & $128\times 64$ & 4 & \\
      & Magnetic rotor & $[-0.5, 0.5]^2$ & $48^2$ & 5 & \ref{fig:srmhd-rotor} \\
    \hline
    \multirow{2}{*}{GRMHD}
      & Black hole accretion: SANE & $[-64 r_g, 64 r_g]^3$ & $96^3$ & 4 & \ref{fig:grmhd-sane} \\
      & Black hole accretion: MAD & $[-1024 r_g, 1024 r_g]^3 $ & $96^3$ & 6\footnote{Limited to $r \leq 100 r_g$, along with the static mesh refinement of level 8 toward the origin.} & \ref{fig:grmhd-mad} \\
    \hline
\end{tabular}
\end{table*}

In this section, we run the AMR simulations for a range of benchmark problems in
Newtonian MHD (Sec.~\ref{sec:newtonian-mhd}) and special relativistic MHD
(Sec.~\ref{sec:results-srmhd}), followed by GRMHD simulations of black hole
accretion disks (Sec.~\ref{sec:3d-grmhd-test}).

All evolved variables are monitored for the grid refinement as in
Algorithm~\ref{indicator-algorithm}, except in the GRMHD accretion tests where
we exclude the energy variable (see Sec.~\ref{sec:3d-grmhd-test}). Across all
the test problems, we use a single pair of the refinement and coarsening
thresholds $[4^{-3}, 4^{-2}]$. The division-by-zero guards for the momentum and
magnetic fields are set to $\epsilon_{\rm mom} = \epsilon_{\rm B} = 10^{-10}$.
This choice is actually substantially lower than the fiducial values of
$(\epsilon_{\rm mom}, \epsilon_B)$ suggested in
Sec.~\ref{sec:amr-adaptation-to-mhd} for all test problems considered here.
Therefore, the results presented in this section show the most sensitive
configuration for the vector variables; adopting the suggested values of
$(\epsilon_{\rm mom}, \epsilon_B)$ would only make the refinement more
selective.

Time evolution is performed with the second order Runge--Kutta time stepper. AMR
criterion (Algorithm~\ref{indicator-algorithm}) is checked every 5 time steps.
For a subset of the tests, we prepare the reference solution run on a uniform
mesh with the grid spacing corresponding to the finest level in the AMR
simulation. Unless specified otherwise, the simulation grid is maximally refined
at $t=0$ in order to minimize the staircasing error from coarsely resolved
features in the initial data; see Sec.~\ref{sec:discussion-initial-refinement}
for the detailed investigations on this caveat. The list of the test problems
and their simulation grid specifications are summarized in
Table~\ref{tab:amr-setup}.

\subsection{Newtonian MHD}
\label{sec:newtonian-mhd}

We use the piecewise parabolic method (PPM4) \cite{Colella:1982ee} for spatial
reconstruction, along with the HLLC approximate Riemann solver
\cite{Harten1983,Einfeldt1991,1994ShWav...4...25T} for hydrodynamic tests and
the HLLD solver \citep{Miyoshi2005} for MHD tests.

\subsubsection{Sod shock tube}

\begin{figure} 
  \centering
  \includegraphics[width=\linewidth]{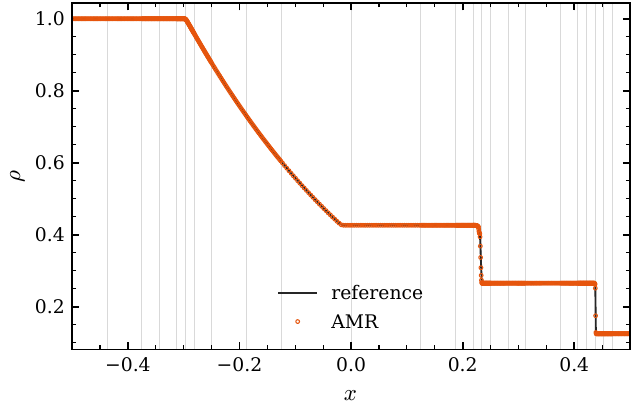}
  \caption{Sod shock-tube test at $t=0.25$. The AMR solution is shown with the
    orange circles, and the reference solution is shown with the black solid
    line. Vertical gray lines mark the boundaries between each
    \texttt{MeshBlock} in the AMR solution.}
  \label{fig:sod}
\end{figure}

The Sod shock tube test \cite{Sod:1978} is a standard problem used to validate
the ability of hydrodynamic codes to accurately capture shocks and contact
discontinuities emerging in the Riemann problems. The fluid is initially at rest
with the density and pressure $(\rho_L, p_L) = (1.0, 1.0)$ on the left side
($x<0$) and $(\rho_R, p_R) = (0.125, 0.1)$ on the right side. Base grid
resolution is $64$ and each \texttt{MeshBlock} has $32$ cells. AMR is allowed up
to level 5, reaching the effective resolution $2048$.

Fig.~\ref{fig:sod} shows the test result at $t=0.25$. The interfaces between the
\texttt{MeshBlock}s in the AMR solution are shown with the gray vertical lines
in the figure, displaying the grid refinement structure. We clearly see that the
mesh refinement is clustered at the head of the rarefaction wave, contact
discontinuity, and shockwave, while remaining on a coarse level on smooth
regions. Compared to the uniform reference solution, the AMR simulation required
$2.85$ times fewer \texttt{MeshBlock}-cycles and completed $1.60$ times faster
in wall-clock time.

\subsubsection{Shu--Osher shock tube}

\begin{figure} 
    \centering
    \includegraphics[width=\linewidth]{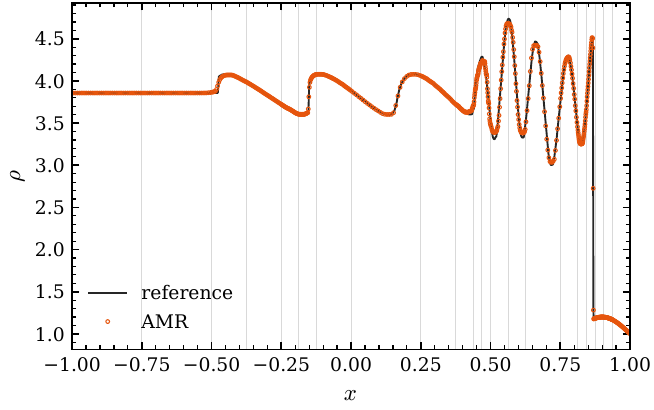}
    \caption{Shu--Osher shock tube test at $t=0.47$, plotted in the same way as
      Fig.~\ref{fig:sod}.}
    \label{fig:shu-osher}
\end{figure}

The Shu--Osher shock tube problem \cite{ShuOsher1989} involves a shockwave
interacting with a smoothly varying background profile, testing the ability of a
code to resolve discontinuities and smooth regions simultaneously. Our setup
used a base grid resolution of $64$ and each \texttt{MeshBlock} has $32$ cells.
AMR is allowed up to level 5, reaching the effective resolution $2048$.

Fig.~\ref{fig:shu-osher} shows the result. The AMR solution correctly captures
the shock position as well as the oscillatory smooth features in the downstream
region. Mesh refinement is concentrated at the shock and the high frequency
waves behind it, whereas the unperturbed region at $x < -0.5$ is held at coarse
resolution. The AMR simulation required $4.26$ times fewer
\texttt{MeshBlock}-cycles than the uniform reference and was $1.72$ times faster
in wall-clock time.

\subsubsection{Kelvin--Helmholtz instability}

\begin{figure} 
    \centering
    \includegraphics[width=\linewidth]{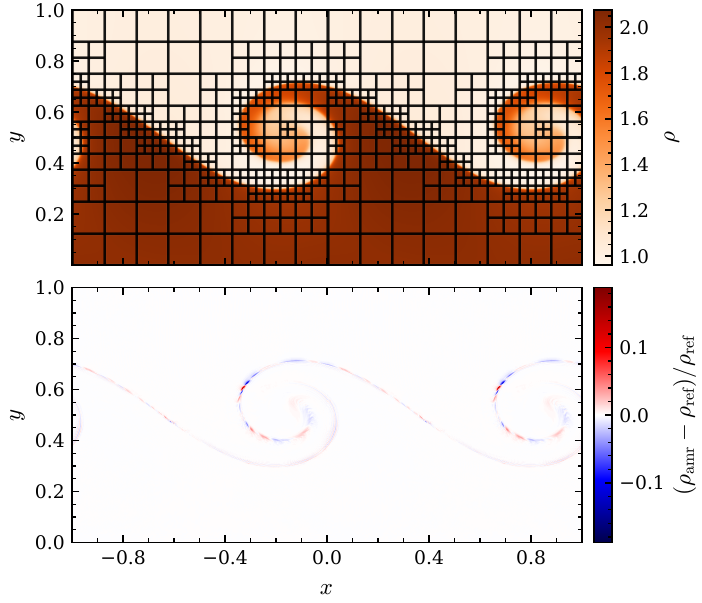}
    \caption{The Kelvin--Helmholtz instability test at $t=1.7$ with PPM4
      reconstruction. {\em Top:} AMR density $\rho$ with \texttt{MeshBlock}
      boundaries. {\em Bottom:} relative error with respect to the reference
      solution. Only the upper half of the full computational domain is shown
      here.}
    \label{fig:kh}
\end{figure}

We use the initial conditions from \citet{Lecoanet_2015} which has a smooth
transition layer between the two shearing fluids. The original initial condition
in \cite{Lecoanet_2015} is shifted to the computational domain $[-1, 1]^2$. We
use the adiabatic index $\Gamma=5/3$. Base grid resolution is $128^2$ and each
\texttt{MeshBlock} has $32^2$ cells. AMR is allowed up to level 4, reaching the
effective resolution $2048^2$.

Fig.~\ref{fig:kh} shows the simulation result at $t=1.7$, when the vortices
between the shear layers have fully developed. Only the upper half $y>0$ is
shown in the figure. The finest mesh structure robustly follows the vortex and
shear boundaries whereas the interior of each layer stays at coarse grid levels.
The error between the AMR and the reference solution is localized to the head of
the vortices, mainly due to the numerical noise at the refinement boundary. The
AMR simulation required $5.26$ times fewer \texttt{MeshBlock}-cycles and was
$1.45$ times faster in wall-clock time than the reference.

\subsubsection{Rayleigh--Taylor instability}

\begin{figure} 
    \centering
    \includegraphics[width=\linewidth]{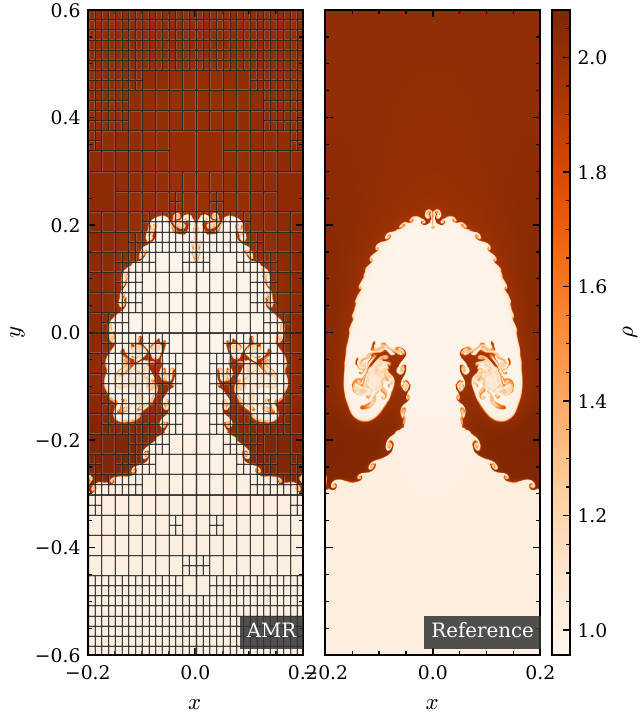}
    \caption{Mass density from the Rayleigh--Taylor instability test at $t=8.5$.
      \texttt{MeshBlock} boundaries are shown for the AMR simulation.}
    \label{fig:rt}
\end{figure}

The initial condition for the Rayleigh--Taylor instability test is drawn from
Ref.~\cite{LiskaWendroff2003}. A constant acceleration $a_g = -0.1$ is added in
the vertical direction. We impose the periodic boundary condition along the
horizontal axis and the reflecting boundary condition along the vertical axis.
The instability is explicitly triggered by adding a sinusoidal velocity
perturbation near the interface at $y=0$. The interface has a smooth hyperbolic
tangent transition layer. Base grid resolution is $32\times64$ and each
\texttt{MeshBlock} has $32^2$ cells. AMR is allowed up to level 5, reaching the
effective resolution $1024\times2048$.

Fig.~\ref{fig:rt} shows the result at $t=8.5$. We see that the AMR solution well
maintains the mirror symmetry about $x=0$, and the finest \texttt{MeshBlock}s
are concentrated along the interface between the fluids. The interior of the
plume is at coarser grid levels.
As the primary instability develops vertically, numerical noise at the interface
gives rise to secondary Kelvin--Helmholtz instabilities. The use of AMR
introduces stronger perturbations for this secondary instability compared to the
unigrid simulation, making the detailed small-scale structures at the interface
differ from the reference solution.
The AMR run was $1.75$ times faster than the reference and required $2.19$ times
fewer \texttt{MeshBlock}-cycles.

\subsubsection{Liska--Wendroff implosion}
\label{sec:test-lwi}

\begin{figure} 
  \centering
  \includegraphics  [width=0.92\linewidth]{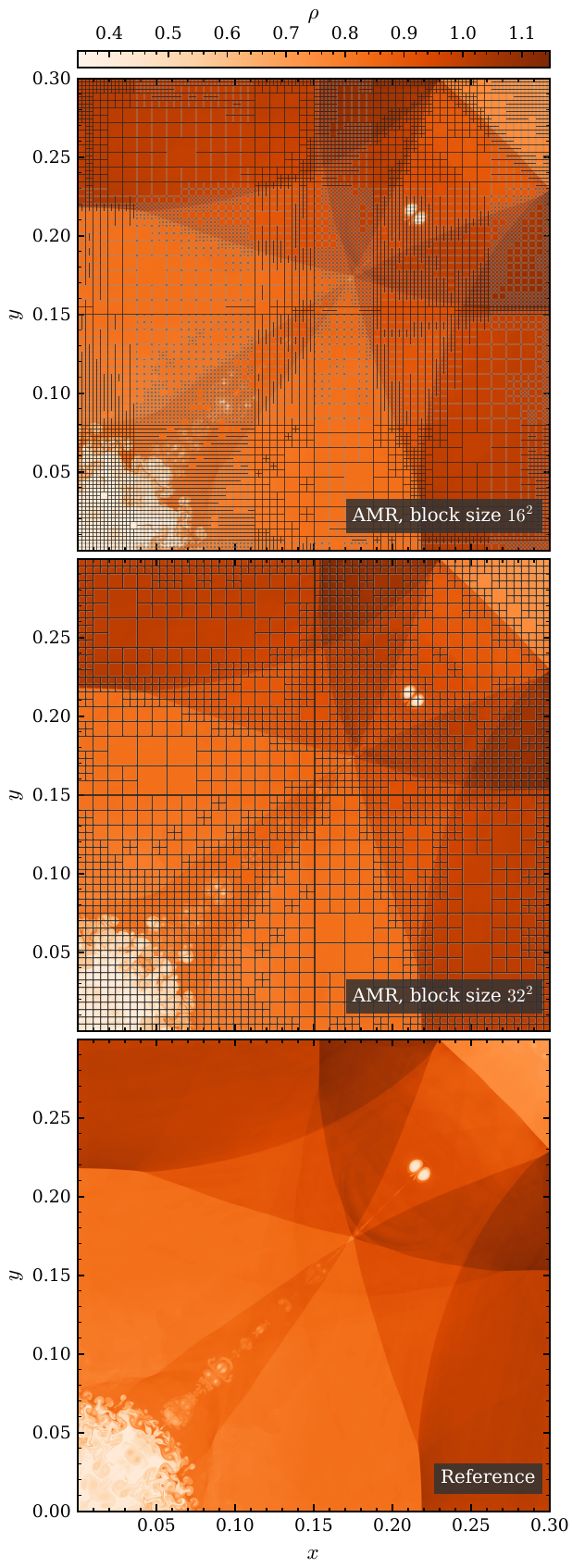}
  \caption{Liska--Wendroff implosion test at $t=1.5$. We show the AMR solution
  using \texttt{MeshBlock} size of $16^2$ (top), $32^2$ (middle) and the unigrid
  reference solution (bottom).}
  \centering
  \label{fig:lwi_3panel}
\end{figure}

The Liska--Wendroff implosion test problem \cite{LiskaWendroff2003} assesses the
symmetry-preserving capability of a code. The domain has the size $[0, 0.3]^2$
and the reflecting boundary condition is imposed on every side. The initial
condition consists of two uniform states of the fluid at rest, divided by a
diagonal $x + y = 0.15$; the density and pressure are $\rho = p = 1$ above, and
$\rho= 0.125$ and $p=0.14$ below. A shock wave is launched into the lower left
corner and reflects back. A repeated passage of the shockwave through the
contact discontinuity triggers the Richtmyer--Meshkov instability and seeds
turbulent vortices along the diagonal $x=y$. The evolution of these vortices is
exquisitely sensitive to whether the diagonal symmetry can be preserved at the
bitwise level. We run two AMR simulations with the same effective resolution of
$2048^2$ but different \texttt{MeshBlock} sizes: $16^2$ grid points per block
and $32^2$ grid points per block, both using a base resolution of $64^2$ and allowed
up to refinement level 5.

Fig.~\ref{fig:lwi_3panel} shows the result from the AMR simulations with two
different \texttt{MeshBlock} sizes and the reference solution. We see that the
refinement is well triggered on the turbulent region in the lower left corner
and the jet developing along the diagonal $x=y$. Smooth regions that are away
from small scale features and shocks remain at intermediate and coarse grid
levels. AMR solutions successfully preserve the exact symmetry, as we can see
from the symmetric mesh refinement structure as well as the turbulent jet lying
along the diagonal. The AMR simulation with \texttt{MeshBlock} size $16^2$ required $2.31$ times fewer
\texttt{MeshBlock}-cycles compared to the reference and was $1.95$ times faster in
wall-clock time.

\subsubsection{Magnetized blast wave}

\begin{figure*}
    \centering
    \includegraphics  [width=\linewidth]{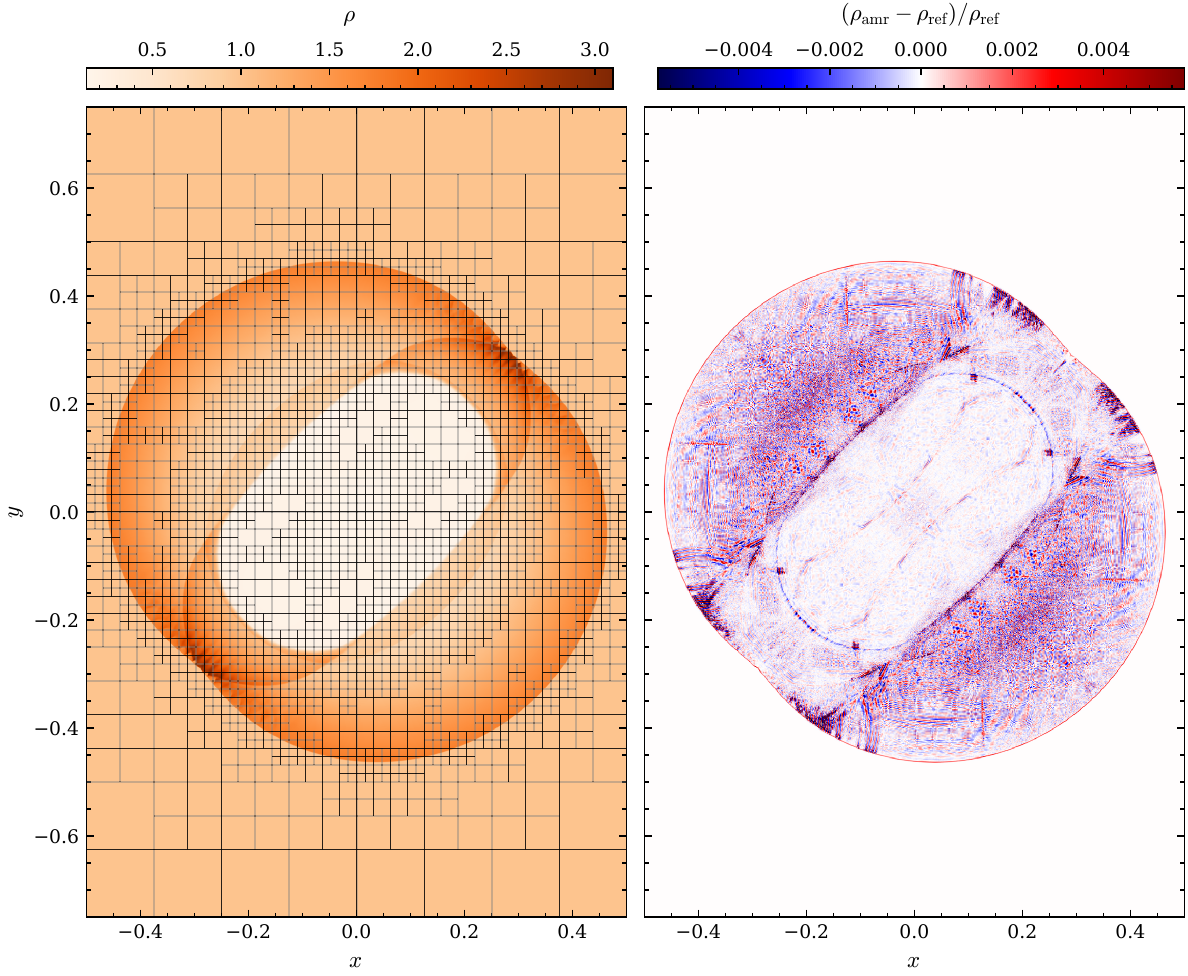}
    \caption{Two-dimensional magnetized blast wave at $t=0.18$. {\em Left:} AMR
      density $\rho$ with \texttt{MeshBlock} boundaries overlaid. {\em Right:}
      relative error with respect to the uniform reference. The finest levels
      track the outer fast shock and the boundary of the low-density cavity,
      while the unshocked ambient medium in the corners of the domain remains
      coarse.}
    \label{fig:mbw}
\end{figure*}

The 2D cylindrical magnetized blast wave problem \cite{Londrillo:1999pm} is set
up following the Refs.~\cite{Gardiner2005,Stone_2008}. The domain has size
$(L_x,L_y)=(1,1.5)$. Initial density and pressure are set to $\rho=1$ and
$p_{\rm out}=0.1$. The inner circular region $r<0.1$ has a higher pressure
$p_{\rm in}=10$. A uniform magnetic field $B^x=B^y=1/\sqrt{2}$ is threaded
everywhere. Base grid resolution is $32\times48$ and each \texttt{MeshBlock} has
$16^2$ cells. AMR is allowed up to level 5, reaching the effective resolution
$1024\times1536$.

Figure~\ref{fig:mbw} shows the AMR solution at $t=0.18$. The low-density cavity
and the dense shell surrounding it are elongated along the magnetic field
direction. The mesh refinement pattern closely follows this structure, where the
unperturbed ambient medium remains on coarse levels. The AMR run required $4.03$
times fewer \texttt{MeshBlock}-cycles than the reference and was $3.66$ times
faster in wall-clock time.

\subsubsection{Orszag--Tang vortex}

\begin{figure*}
    \captionsetup[subfloat]{captionskip=1em}
    \centering
    \subfloat[\label{fig:otv-sub1} Mass density and AMR grid level at $t=0.24$
      (left), $t=0.5$ (middle) and $t=0.76$ (right). The rectangle on the upper
      right panel corresponds to the region shown in Fig.~\ref{fig:otv-sub2}. ]{
      \includegraphics[width=0.98\linewidth]{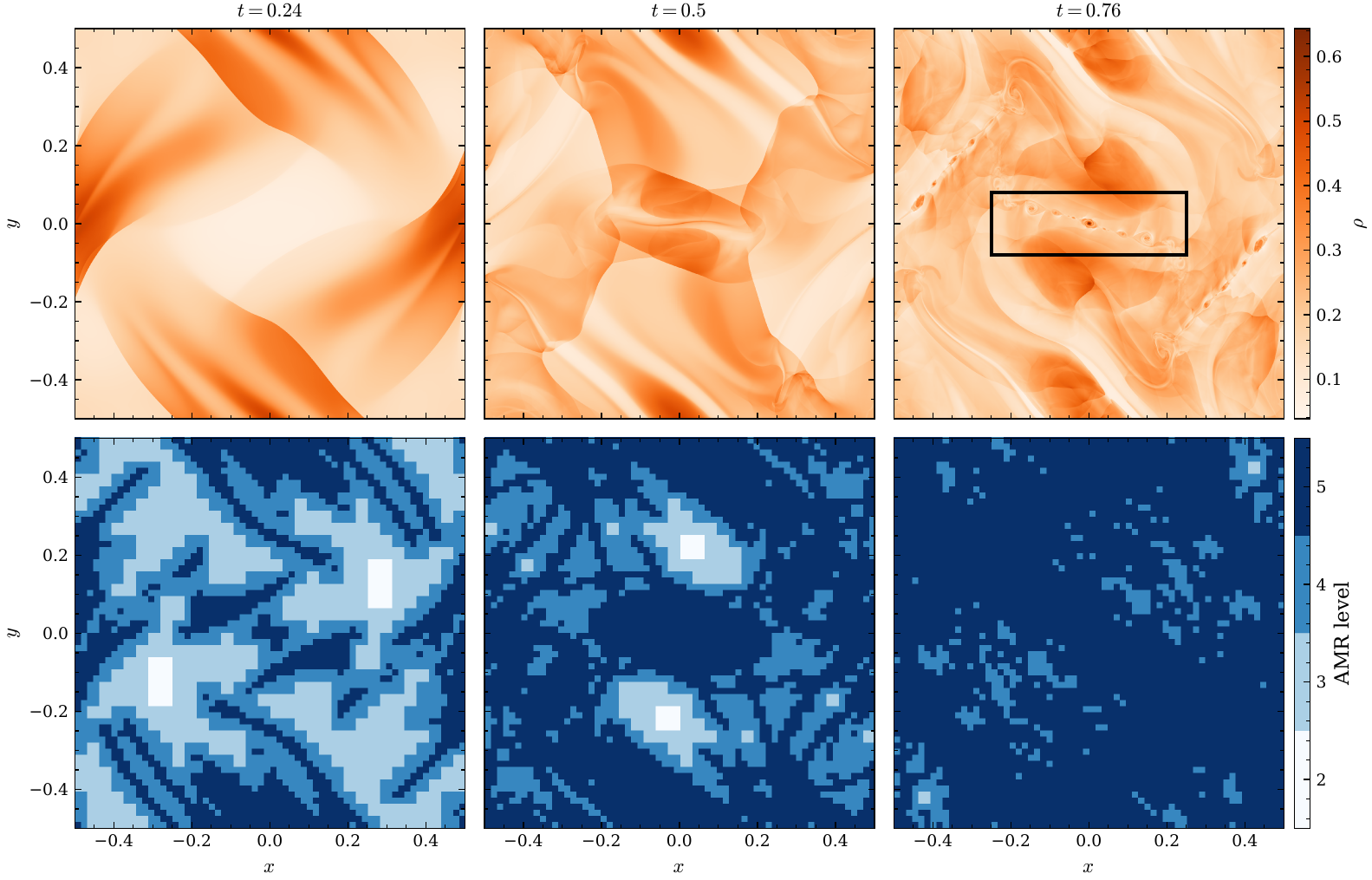}
    }\\[1em]
    \subfloat[\label{fig:otv-sub2} Zoom-in of the central reconnection layer at
      $t=0.76$. Shown are the mass density (top), magnetic field magnitude and
      field lines (middle), and the AMR grid level (bottom). Left column is the
      zoom-in of the result shown in Fig.~\ref{fig:otv-sub1}, and the right
      column is from a separately performed AMR simulation that does not include
      the magnetic field magnitude in the refinement criterion.]{
      \includegraphics[width=0.98\linewidth]{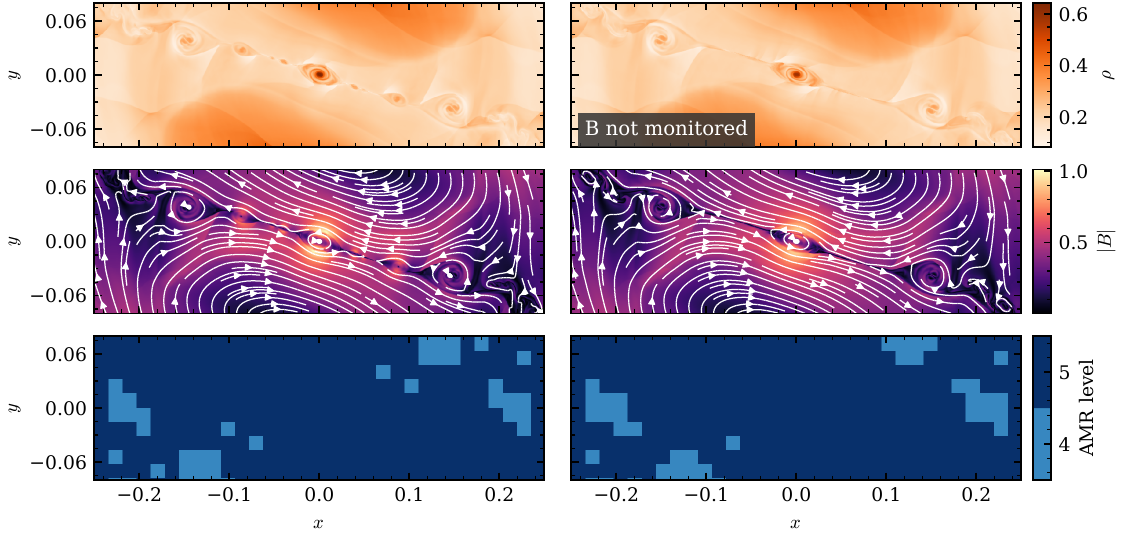}
    }
    \caption{Orszag--Tang vortex problem.}
    \label{fig:ot}
\end{figure*}

The Orszag--Tang vortex \cite{1979JFM....90..129O} is a widely used benchmark
problem for validating multi-dimensional MHD solvers. We use the same setup used
in Ref.~\cite{Stone_2008}. Computational domain has the size $[-0.5, 0.5]^2$
with the periodic boundary condition imposed. Initially the fluid has a constant
density $\rho= 25/(36\pi)$ and fluid pressure $p = 5/(12\pi)$, with the velocity
$v^i = (-\sin 2\pi y, \sin 2 \pi x)$ and the magnetic field $B^i = (-\sin 2\pi
y, \sin 4 \pi x) / \sqrt{4\pi}$. We use the adiabatic index $\Gamma = 5/3$.
Starting from a smooth initial condition, the vortical initial velocity field
drives interactions between MHD shocks and the flow becomes supersonically
turbulent in later times. Base grid resolution is $64^2$ and each
\texttt{MeshBlock} has $16^2$ cells. AMR is allowed up to level 5, reaching the
effective resolution $2048^2$.

Fig.~\ref{fig:otv-sub1} shows the simulation result. The AMR refinement is
robustly triggered along the shocks, where smooth plateaus between the shocks
remain on coarse grid levels. The fraction of the refined grid levels in the
computational domain increases over time as the flow becomes more turbulent.
Consequently, the gain from AMR diminishes over the course of the evolution.
Over the full run to $t=2.0$, the AMR simulation required only $1.18$ times
fewer \texttt{MeshBlock}-cycles than the reference and was $1.08$ times faster
in wall-clock time, with nearly all of the savings accumulated at early times
before the shocks fill the domain.

In Fig.~\ref{fig:otv-sub2}, we compare the result with a separate simulation
that does not include the magnetic field magnitude in the AMR criterion. While
the grid is mostly refined to the maximum level within the zoomed-in rectangular
region for both runs, the original AMR simulation that monitors the magnetic
field magnitude better resolves the secondary plasmoids lying along the
reconnection layer. This comparison in particular demonstrates that properly
resolving small-scale features in AMR simulations may require not only capturing
them once they have fully emerged in the numerical solution, but also an early
detection of such features in their developing stage.

\subsection{SRMHD}
\label{sec:results-srmhd}

In the SRMHD tests, we use the PPM4 reconstruction and the HLLE Riemann fluxes
for the spatial discretization. The first-order flux correction
\cite{Lemaster:2008gh,Stone:2020,Stone_2026} is switched on. The AMR indicator
directly monitors the conserved variables of the SRMHD solver in
\athenak~\cite[see][]{Stone_2026}.

\subsubsection{Magnetized blast wave}
\label{sec:results-srmhd-blast}

\begin{figure}
    \centering
    \includegraphics[width=\linewidth]{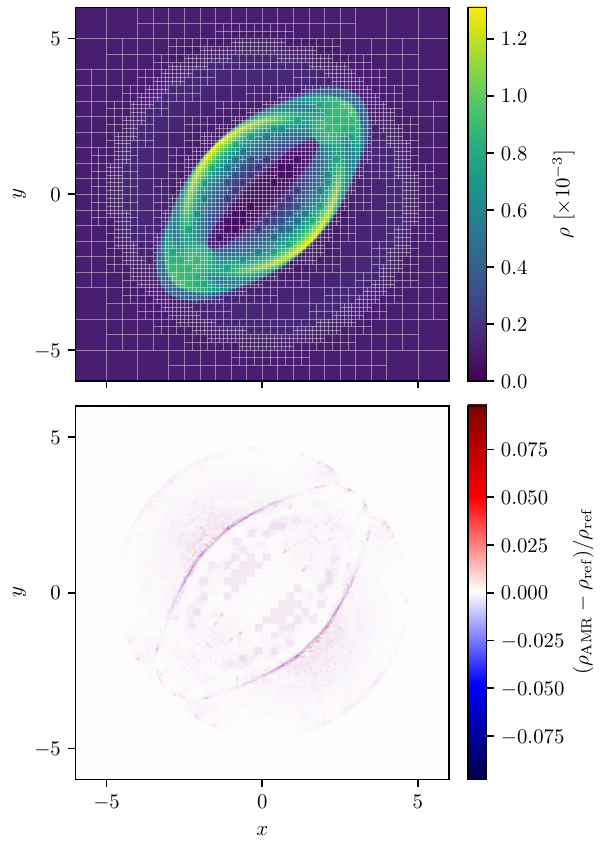}
    \caption{2D relativistic magnetized blast wave at $t = 4.0$. {\em Top:}
      rest-mass density $\rho$ with the mesh boundaries overlaid as white solid
      lines. Each square block corresponds to a \texttt{MeshBlock} containing
      $16^2$ cells. {\em Bottom:} relative difference between the AMR and the
      reference solution.}
    \label{fig:srmhd-blast-2d}
\end{figure}

We follow the initial condition for the 2D relativistic magnetized blast wave
test used in Ref.~\cite{Stone:2020}. Density and pressure are set to $\rho =
10^{-2}$ and $p = 1$ within $r \leq 0.8$, and $\rho = 10^{-4}$ and $p = 5\e{-3}$
for $r \geq 1$. Both quantities linearly vary in the transition region $0.8 < r
< 1$. The gas is initially at rest, $v^i = 0$, and threaded by a uniform
magnetic field $B^x = B^y = 1/\sqrt{20}$ and $B^z = 0$. We use the adiabatic
index $\Gamma = 4/3$. Base grid resolution is $48^2$ and each \texttt{MeshBlock}
has $16^2$ cells. The maximum AMR level is 5, reaching the finest grid
resolution equivalent to the reference resolution $1536^2$.

Fig.~\ref{fig:srmhd-blast-2d} shows the test result. The finest grid level
faithfully tracks the discontinuous features, while the unshocked ambient
regions and smooth part of the solution remain at the coarse grid level. The
number of \texttt{MeshBlock}s is about 40\% by the end of the simulation
($t=4.0$) and the overall speedup is $2.2\times$ compared to the unigrid
reference simulation.

\subsubsection{Magnetized Kelvin--Helmholtz instability}
\label{sec:results-srmhd-khi}

\begin{figure}
    \centering
    \includegraphics[width=\linewidth]{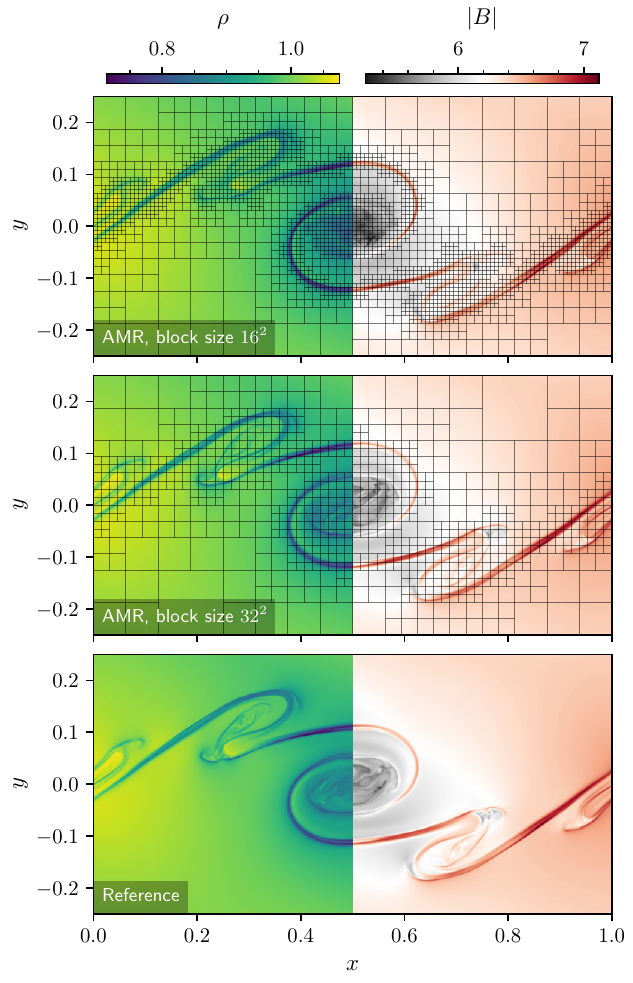}
    \caption{Relativistic magnetized Kelvin--Helmholtz instability test at $t =
      5.0$. The rest-mass density (left) and the magnetic field magnitude
      (right) are shown on each half of the computational domain. We show the
      AMR solution using the MeshBlock size of $16^2$ (top), $32^2$ (middle),
      and the unigrid reference solution (bottom).}
    \label{fig:srmhd-khi}
\end{figure}

The magnetized Kelvin--Helmholtz instability problem is initialized following
Refs.~\cite{Mignone:2011fd,Stone:2020}. The initial state has uniform density
$\rho = 1$ and pressure $p = 20$, with magnetic field $B^x = \sqrt{2/5}$, $B^y =
0$, and $B^z = 10\sqrt{2/5}$. The initial velocity is
\begin{align}
  v^x &= \frac{1}{4} \tanh (100y), \\[1ex] 
  v^y &= \frac{1}{400} \sin (2\pi x) \exp(-100y^2), \\[1ex]
  v^z &= 0 .
\end{align}
The adiabatic index is set to $\Gamma = 4/3$. Periodic boundary condition is
imposed in the $x$ direction and the outflow boundary is used in $y$ direction.
We run two AMR simulations with the same effective resolution of
$2048\times1024$ but different \texttt{MeshBlock} sizes: $16^2$ grid points per
block allowing the refinement level 5 (base grid $64\times 32$) and $32^2$ grid
points per block allowing up to the level 4 (base grid $128\times 64$).

Figure~\ref{fig:srmhd-khi} shows the test result at $t = 5.0$. Overall, AMR
solutions accurately track filamentary structure and vortices developing at the
fluid interface, while keeping the smooth regions at coarse grid levels. We see
that the AMR run using a smaller block size (therefore more crinkly mesh
structure) is subject to a higher level of refinement noise, altering the
turbulent structure within the primary vortex at the center and slightly
shifting the position of the interface. The second AMR run using the block size
of $32^2$ shows a significant improvement. Compared to the uniform grid
reference simulation run with the same \texttt{MeshBlock} size, the $32^2$ AMR
simulation showed 22\% of the total \texttt{MeshBlock}-cycles and 25\% of the
wall-clock time.

\subsubsection{Magnetic rotor}
\label{sec:results-srmhd-rotor}

\begin{figure*}
    \centering
    \includegraphics[width=\linewidth]{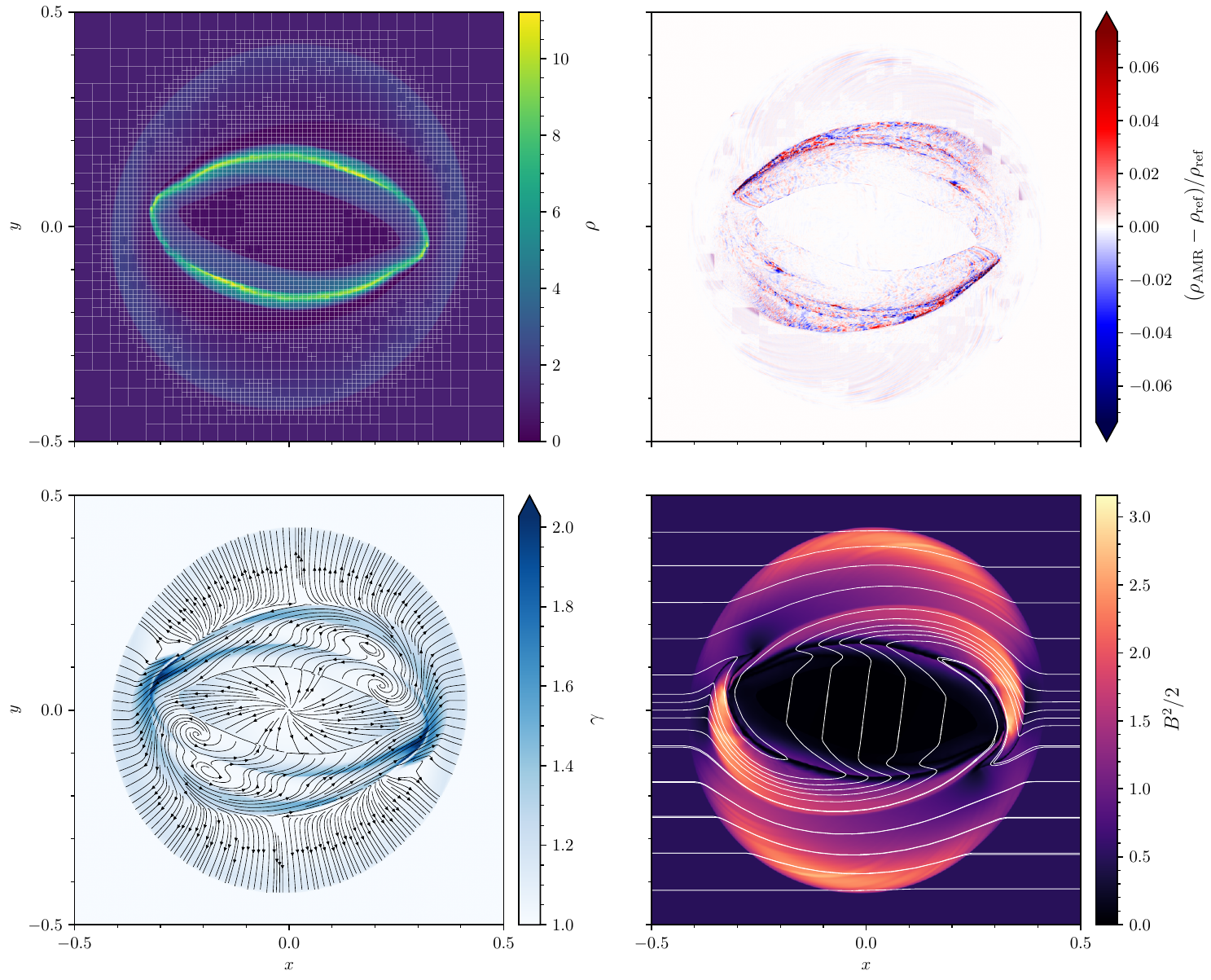}
    \caption{Magnetic rotor test at $t=0.4$, showing the rest mass density with
      the mesh structure (top left), relative error in the rest mass density
      with the reference solution (top right), fluid Lorentz factor with the
      velocity field lines (bottom left), and magnetic pressure with the
      magnetic field lines (bottom right).}
    \label{fig:srmhd-rotor}
\end{figure*}

In the 2D magnetic rotor problem \cite{BALSARA1999270,DelZanna:2002rv}, a
rotating disk is placed inside a less dense uniform fluid. Inside the disk
$r\leq 0.1$, the initial density is $\rho=10$ and the velocity is $v^i = 9.95
(-y, x, 0)$, surrounded by the ambient region at rest with $\rho = p = 1$. A
uniform magnetic field $B^x = 1$ is threaded everywhere. We use the adiabatic
index $\Gamma=5/3$. Base grid resolution is $48^2$ and each \texttt{MeshBlock}
has $16^2$ cells. AMR is allowed up to level 5, reaching the effective
resolution $1536^2$.

Fig.~\ref{fig:srmhd-rotor} shows the result. Magnetic braking slows down the
rotor and the rotation angle reaches approximately 90 degrees by the final
simulation time $t=0.4$. A somewhat extended coverage of the refined region
surrounding the oval overdense structure, seemingly smooth in terms of the rest
mass density, is mostly triggered from the slow falloff of the magnitude of the
momentum variable. The residual error between the AMR and the reference solution
is concentrated to the elongated tips of the oval overdense region. The AMR
simulation shows 3.4$\times$ saving in the \texttt{MeshBlock}-cycles and
2.8$\times$ speedup.

\subsection{GRMHD: Black hole accretion}
\label{sec:3d-grmhd-test}

We simulate the black hole accretion disks in the standard and normal evolution
(SANE) \cite{Narayan:2012yp,Sadowski:2013joa} and magnetically arrested disk
(MAD) \cite{Narayan:2003by,Tchekhovskoy:2011zx} regimes, showcasing the
potential of our AMR criteria in large-scale GRMHD simulations. Simulations are
performed in the scale-free manner in which the simulation coordinates and time
are normalized with $r_g \equiv GM_{\rm BH}/c^2$ and $r_g c^{-1} = GM_{\rm
BH}/c^3$.

For both simulations, a Fishbone--Moncrief torus \cite{Fishbone:1976} is
initialized in the Kerr spacetime with the black hole spin $a/M_{\rm BH} = 0.9$.
Random perturbations at the 2\% level in the internal energy are injected to
trigger the magneto-rotational instability (MRI) \citep{Balbus:1991ay}. The
torus is initially embedded in the background rest-mass density $\rho_{\rm
bg}(r) = 10^{-5} r^{-1.5}$ and the pressure $p_{\rm bg}(r) = \rho_{\rm bg}(r) /
(300r)$. We use the piecewise linear reconstruction and the HLLE Riemann fluxes
for spatial discretization. The fluid Lorentz factor is capped to $W_{\rm max} =
20$ and the first-order flux correction is switched on.

The GRMHD solver of \athenak~for a non-dynamical spacetime employs the `HARM
trick' \cite{Gammie:2003rj,Stone_2026} that subtracts the rest energy density
contribution from the energy equation. As a result, the evolved variable of the
energy equation, $\sqrt{-g}(\tensor{T}{^t_t} + \rho u^t)$, in general does not
have a fixed sign in a curved spacetime. This zero-crossing behavior can be
problematic for the AMR indicator as noted earlier. In such cases, one can
instead monitor a corresponding primitive variable (e.g. specific internal
energy) that is strictly positive. For the GRMHD tests performed in this
section, however, we simply do not monitor the conserved variable of the energy
equation.

\subsubsection{SANE}
\label{sec:grmhd-sane}

\begin{figure*}
    \centering
    \includegraphics[width=\linewidth]{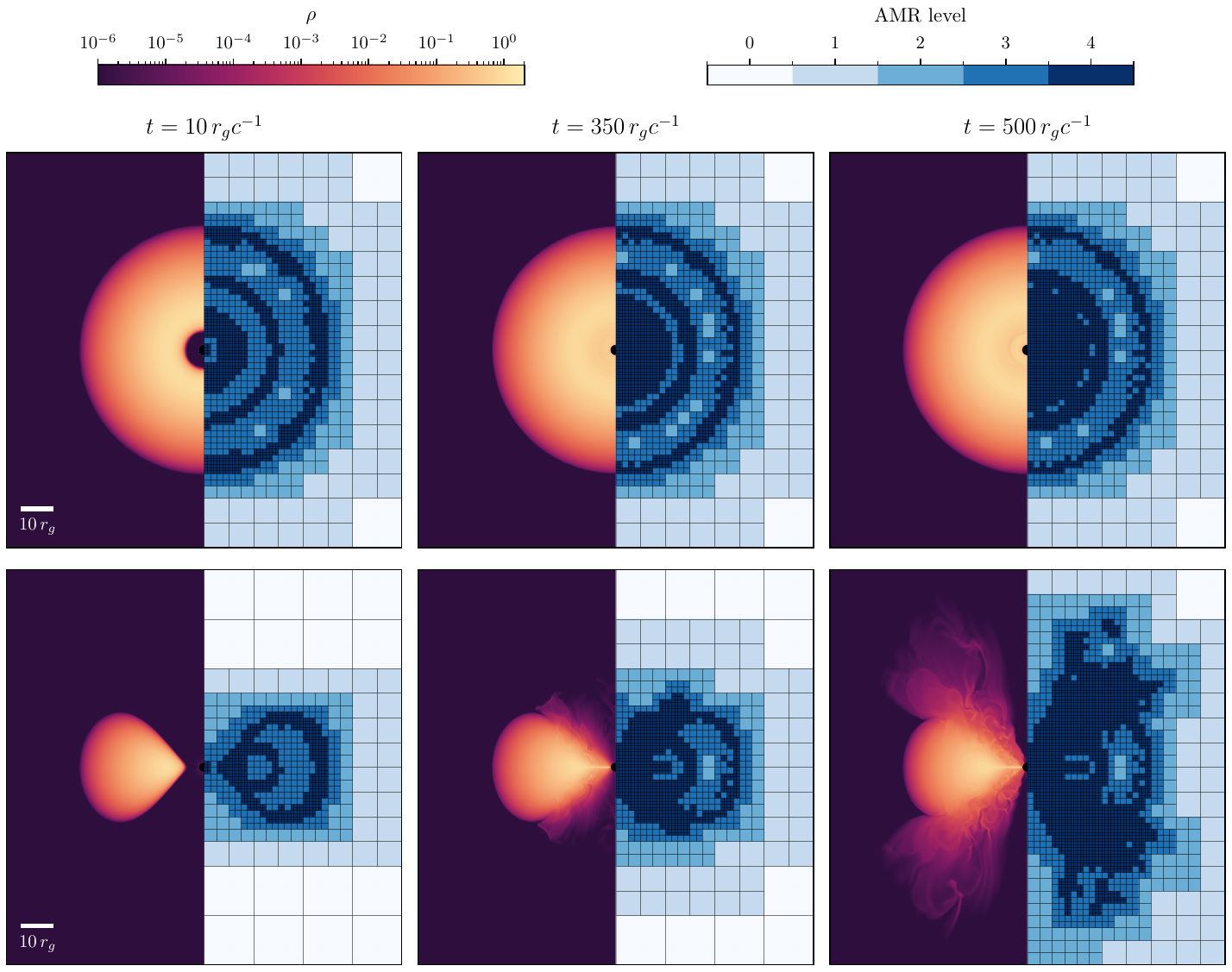}
    \caption{AMR simulation of the SANE accretion disk. Shown are the equatorial
    slice (top row) and vertical slice (bottom row) for different simulation
    times (left to right columns). We show the rest mass density and the AMR
    grid structure in each half in the panels.}
    \label{fig:grmhd-sane}
\end{figure*}

The initial torus has its inner edge at $r_{\rm in} = 6r_g$ and the outer edge
at $r_{\rm out} = 41 r_g$, with the pressure maximum located at $r_{\rm peak} =
12 r_g$. Magnetic field is initialized using the following form of the vector
potential
\citep{Wong:2021qbe,Wong:2022rqr}
\begin{equation} \label{eq:grmhd-initialB-sane}
  A_\phi = \max\left[ \rho - 0.2, 0 \right] \, .
\end{equation}
The strength of the magnetic field is then globally rescaled so that the ratio
between the maximum fluid pressure $p$ and the maximum comoving magnetic
pressure $p_{\rm mag} = b^\mu b_\mu/2$ within the torus equals $p / p_{\rm mag}
= 100$. We use the ideal gas equation of state with $\Gamma = 4/3$. The density
and pressure atmosphere values are set to $(\rho_{\rm floor}, p_{\rm floor}) =
(10^{-10}, 3\e{-12})$ for the time evolution.

The computational domain spans $[-64r_g, 64r_g]^3$ with the base grid resolution
$96^3$. Each \texttt{MeshBlock} has $12^3$ cells, and up to four levels of
additional refinement are allowed, reaching the finest resolution of 12 grid
points per $r_g$. The innermost region $[-2r_g, 2r_g]^3$ containing the black
hole always stays at the maximum refinement level. Initially, the whole torus is
enclosed with two levels of mesh refinement relative to the coarsest grid level.
We use the floor $D_{\rm floor} = 10^{-5}$ in terms of the evolved variable $D =
\sqrt{-g} \rho u^t$ below which the refinement criterion is skipped.

Fig.~\ref{fig:grmhd-sane} shows the rest-mass density and the AMR grid structure
on the equatorial and vertical slices at different simulation times. Soon after
the simulation starts ($t=10 r_gc^{-1}$), mesh refinement is concentrated along
the surface of the torus. As MRI begins to develop from the inner edge of the
torus and accretion onto the black hole begins ($t=350 r_gc^{-1}$), mesh
refinement is triggered following the turbulent inner accretion flow whereas the
outer region in the torus remains at coarser grid levels. As MRI fully develops
and high-latitude outflow is launched ($t=500 r_gc^{-1}$), progressively a
higher fraction of the computational domain is filled with turbulent flows
tracked by fine grid levels. The total number of \texttt{MeshBlock}s at $t=500
r_gc^{-1}$ is about $3.4\e{5}$.

\subsubsection{MAD}
\label{sec:grmhd-mad}

\begin{figure*}
    \centering
    \includegraphics[width=0.99\linewidth]{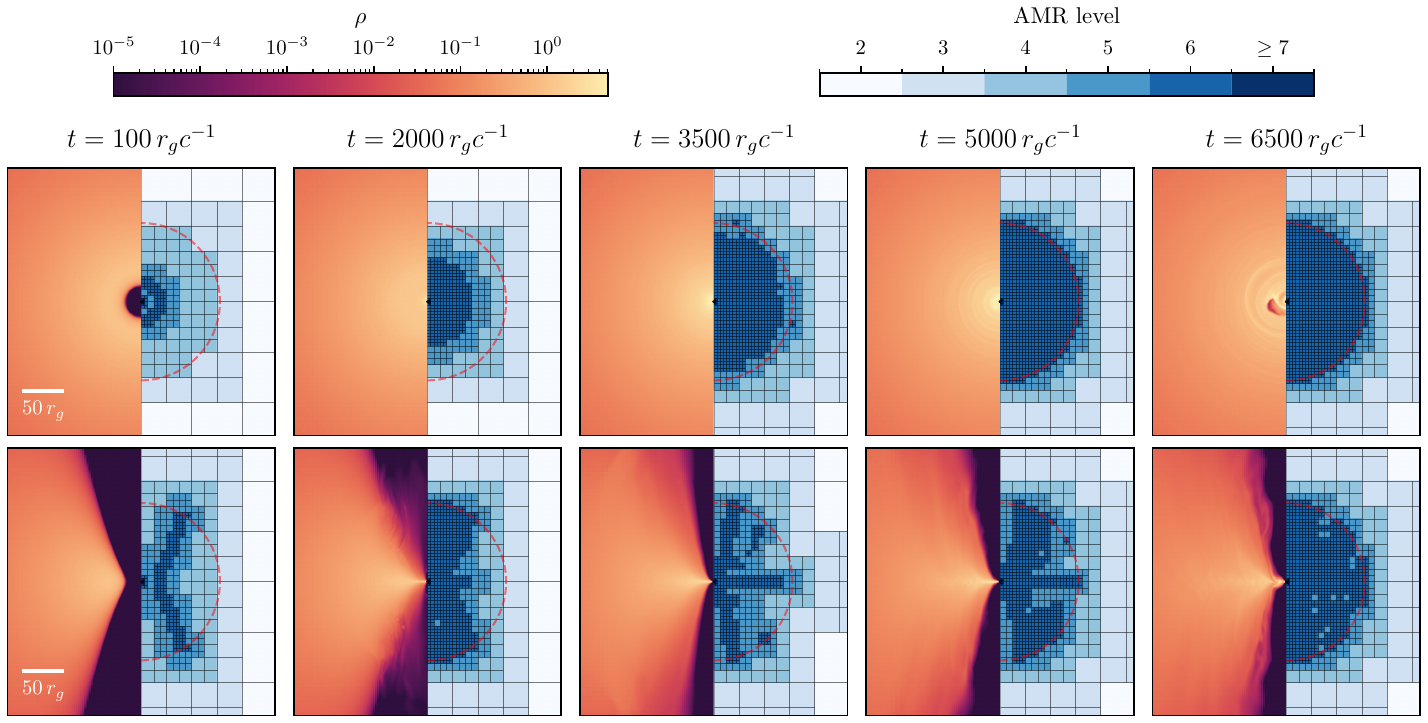}
    \caption{AMR simulation of the MAD accretion disk. Rest mass density (left)
      and AMR grid structure (right) on the equatorial and vertical planes at
      different simulation times. Grid refinement up to level 6 is allowed
      within the radial mask $r \leq 100 r_g$, shown as the dashed red line.}
    \label{fig:grmhd-mad}
\end{figure*}

For the MAD accretion disk, the magnetic field is initialized with
\begin{equation} \label{eq:grmhd-initialB-mad}
    A_\phi = \max\left[ \rho
      \left(\frac{r \sin \theta}{r_{\rm in}}\right)^3
      e^{-r/400} - 0.2, 0 \right] \, ,
\end{equation}
and again rescaled to have $\max(p) / \max(p_{\rm mag}) = 100$ in the torus. The
initial torus has the inner edge at $r_{\rm in} = 20r_g$, pressure maximum at
$r_{\rm peak} = 41r_g$, and the outer edge at $r_{\rm out} = 900r_g$. Compared
to the SANE disk, Eq.~\eqref{eq:grmhd-initialB-mad} concentrates the initial
magnetic field more toward the inner edge of the disk
\citep{Wong:2021qbe,Wong:2022rqr}. We use the ideal gas equation of state with
$\Gamma = 13/9$. The floor values for the rest mass density and pressure are set
to $(\rho_{\rm floor}, p_{\rm floor}) = (10^{-8}, 3\e{-10})$ for the GRMHD
solver.

The computational domain spans $[-1024r_g, 1024r_g]^3$. The base grid resolution
is $96^3$ and each \texttt{MeshBlock} has $12^3$ cells. The innermost region
$[-4r_g, 4r_g]^3$ containing the black hole is fixed to stay at the grid
refinement level 8, having 12 grid points per $r_g$. In order to make this
large-scale test simulation computationally affordable, our AMR criterion is
allowed to refine up to the refinement level 6 (i.e. 3 grid pts/$r_g$) only
within $r \leq 100r_g$. Initially, the region $[-100 r_g, 100 r_g]^3$ is
resolved at the grid refinement level 6 to reduce the staircasing artifact along
the torus surface. Similar to the SANE disk simulation, $D_{\rm floor} =
10^{-5}$ is set as a density floor for the mesh refinement.

In Fig.~\ref{fig:grmhd-mad}, we show the time evolution of the rest mass density
and the AMR grid levels on the equatorial ($xy$) and vertical ($yz$) planes.
Similar to the SANE simulation, grid refinement is confined to the surface of
the torus in early times. The onset of MRI drives a transient launching of
magnetized wind ($t=2000 r_g c^{-1}$ column in the Figure) and the mesh
refinement is temporarily concentrated to the bipolar region through which the
wind is launched. After the wind launching, the mesh refinement stays clustered
on the equatorial plane and the jet--disk boundary ($t=3500 r_g c^{-1}$ column
in the Figure). By $t \approx 5000 r_g c^{-1}$, the dimensionless magnetic flux
on the black hole horizon reaches $\phi_{\rm BH} = \Phi_{\rm BH}/\sqrt{\dot{M}}
\gtrsim 15$ and the accretion enters the MAD regime, featuring a strongly
magnetized jet launched along the spin axis of the black hole
\cite{Blandford:1977ds,Tchekhovskoy:2011zx}. The refined region along the
jet--disk boundary expands over time in response to the highly fluctuating and
turbulent flow structure. Following the first flux eruption event at $t\approx
6000 r_g c^{-1}$, we see that the entire mask $r\leq 100r_g$ is filled with the
finest grid level. In a similar vein to the Orszag--Tang vortex test, the MAD
accretion test shows that there is no significant cost saving from AMR once the
flow becomes fully turbulent within the region of interest.

\section{Conclusion and remarks}
\label{sec:discussion}

\begin{figure*}
    \centering
    \includegraphics[width=\linewidth]{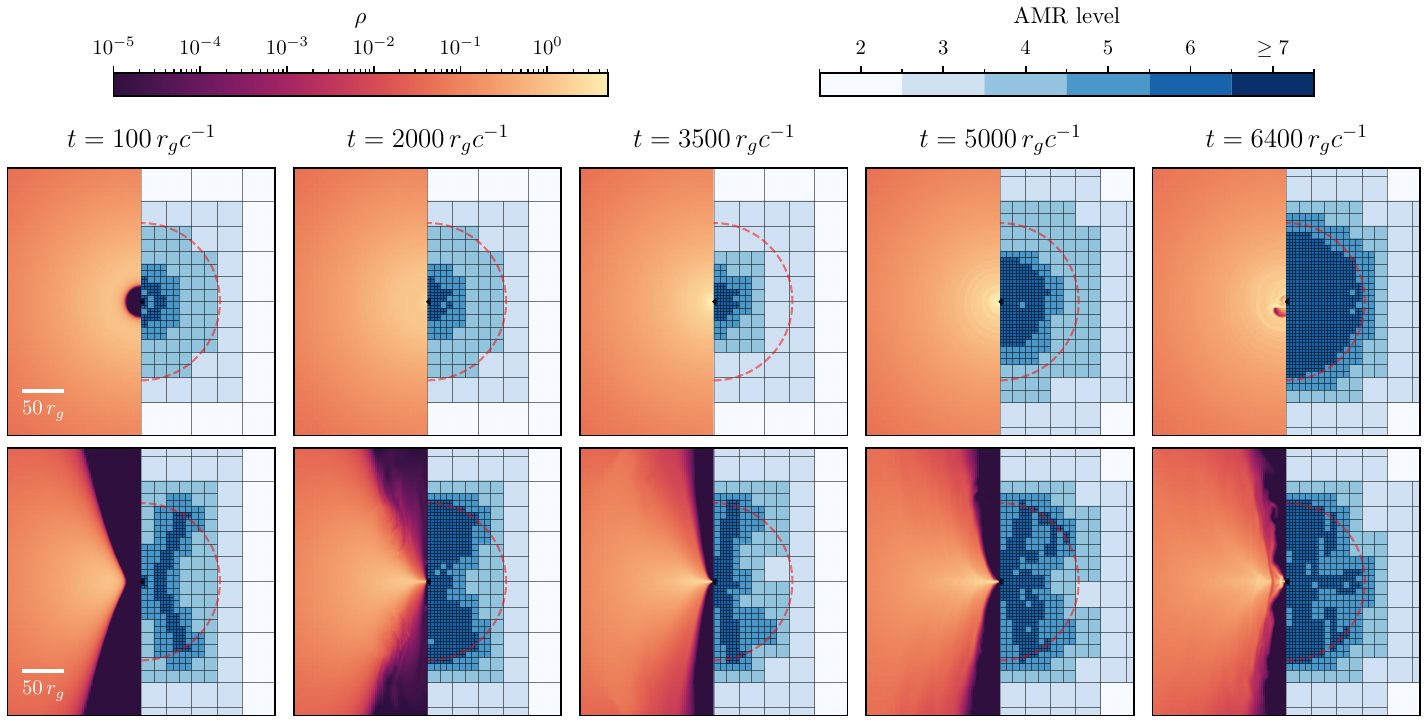}
    \caption{AMR simulation of the MAD accretion disk using the identical setup
      as Fig.~\ref{fig:grmhd-mad} but only monitoring the conserved density
      variable $D = \sqrt{-g} \rho u^t$.}
    \label{fig:grmhd-mad-densonly}
\end{figure*}

\begin{figure*}
\centering
  \includegraphics[width=\linewidth]{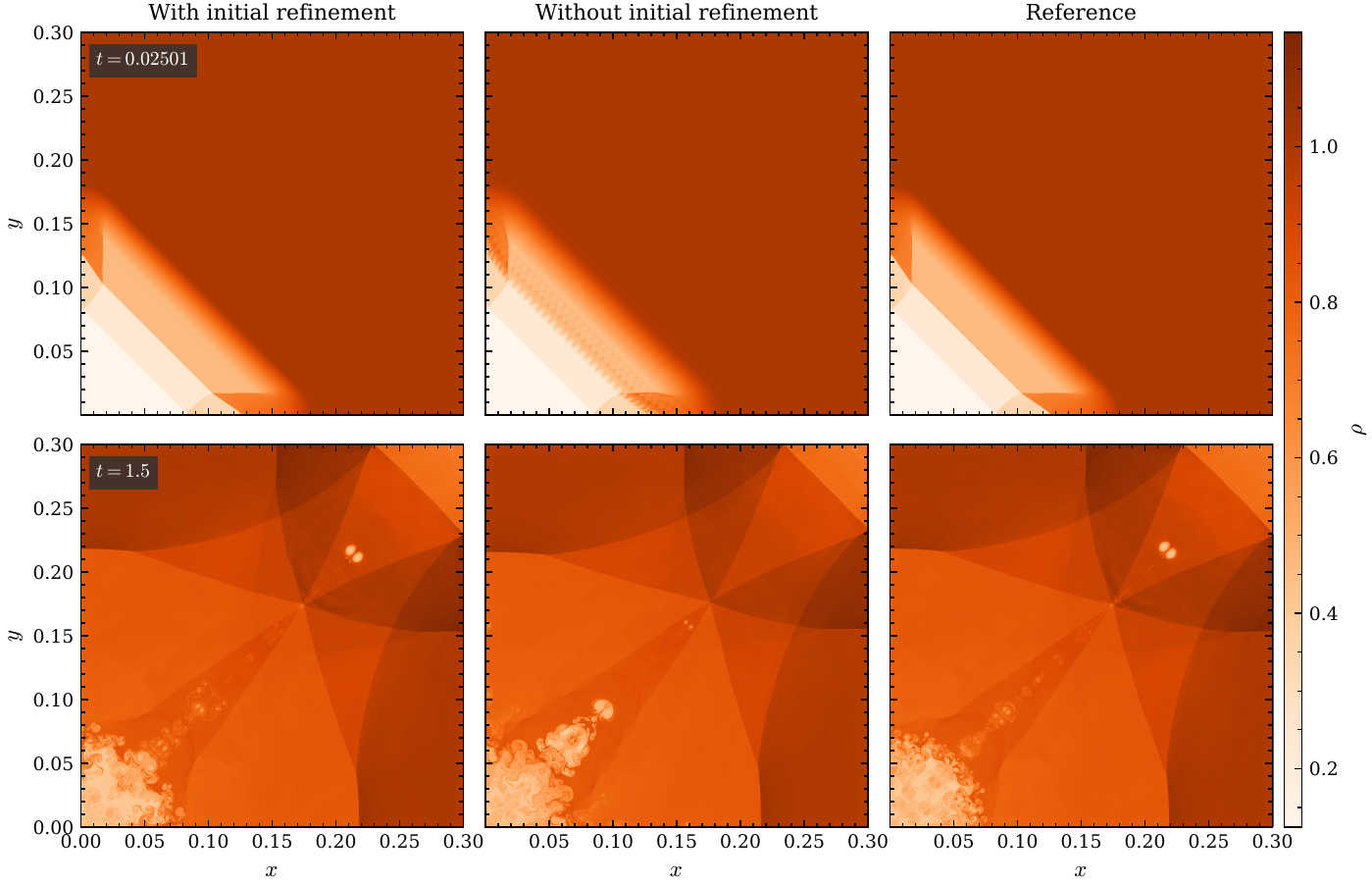}
  \caption{Fluid mass density in the Liska--Wendroff implosion problem at
    $t=0.025$ (top) and $t=1.5$ (bottom). Shown are the AMR simulation with a
    full initial refinement of the domain (left), AMR without initial refinement
    (middle), and the unigrid reference solution (right), all with the same
    finest grid resolution.}
  \label{fig:lwi_noInitRef}
\end{figure*}

\subsection{Summary and outlook}

In this work, we have developed a new AMR criterion targeted for MHD simulations
and have tested its applicability over a range of benchmark problems in
Newtonian and relativistic MHD. Our criterion borrows its idea from the shock
indicators used in the spectral methods, monitoring the modal falloff of the
numerical solution over a 5-point wide finite volume stencil. The scale-free
nature of this AMR criterion, similar to the conventional L{\"o}hner method
\cite{1987CMAME..61..323L}, is well suited to the simulations with fully
adaptive mesh refinement.

Our newly proposed criterion responds to the local behavior of the solution
rather than to a prescribed threshold or numerical slope of physical variables.
Adopting a fixed, single pair of the refinement and coarsening thresholds
$[4^{-3}, 4^{-2}]$ across all the benchmark problems tested, our criterion
robustly triggers the mesh refinement at shocks, contact discontinuities, shear
layers, and current sheets, while holding smooth regions and unperturbed ambient
media at coarse levels.
Augmented with extra algorithmic guards to limit the spatial region or the
regime of physical interest, the new criterion is anticipated to be applicable
to large-scale MHD simulations where fully dynamic AMR might be computationally beneficial. 

While we have focused on the Newtonian and relativistic ideal MHD simulations in
this paper, the applicability of our criterion should not be limited to this
scope. A combined usage with existing criteria that are empirically well suited
for a specific code framework or a problem under consideration, is also
certainly possible.

\subsection{Choice of monitored variables}
\label{sec:discussion-variables}

For the test simulations in this paper, we have monitored the maximal set of the
conserved (evolved) variables, prioritizing robustness over optimizing
computational cost. Empirically, we find many test problems can be resolved in a
reasonable manner by only monitoring the evolved variable associated with the
(rest-) mass density. Still, problems including shear layers, such as the
Kelvin--Helmholtz instability, benefit substantially from including the momentum
variables in the AMR indicator. Inclusion of the magnetic field magnitude
provides an improved sensitivity to the current sheets, which might not
necessarily spatially coincide with the hydrodynamic features that trigger the
mesh refinement. The specific choice on the list of the variables monitored
(e.g. in lines 8-11 of Algorithm~\ref{indicator-algorithm}) might depend on the
problem of interest, while we have demonstrated the most comprehensive case in
this work.

As an instructive example, we perform an additional simulation of the MAD
accretion disk (Sec.~\ref{sec:grmhd-mad}) only monitoring the evolved variable
$D = \sqrt{-g} \rho u^t$ associated with the rest mass density for AMR, and show
the results in Fig.~\ref{fig:grmhd-mad-densonly}. Compared to monitoring all
evolved variables except the energy variable (Fig.~\ref{fig:grmhd-mad}), the
fraction of the refined region decreases, but not so dramatically. We see that
the equatorial plane begins to be refined at a relatively later time, and a
fraction of the spherical region $r\leq 100 r_g$ remains at one level coarser
still after the first flux eruption event. The overall trend in the time
variation of the mesh refinement, however, is quite similar to the previous
case.

\subsection{Initial grid refinement}
\label{sec:discussion-initial-refinement}

In this work we have initialized the AMR simulations with the entire domain at
the maximum refinement level if feasible (except the black hole accretion
simulations, which require a prohibitively large amount of memory), to ensure
that the initial conditions are adequately resolved. This procedure is
especially important for a benchmark problem that is sensitive to the initial
condition or the level of numerical dissipation at the earliest phase of
evolution. Otherwise, the residual between the AMR solution and the uniform
reference solution contains the amplified contribution from an unequal
\emph{initial discretization}, obscuring the assessment of the error solely
coming from AMR itself.

Figure~\ref{fig:lwi_noInitRef} shows the impact of the initial grid refinement
in the Liska--Wendroff implosion test. The failure to resolve the discontinuity
in the initial data at the identical finest resolution results in significant
staircasing artifacts (first row in the Fig.~\ref{fig:lwi_noInitRef}) that
affect the solution in later times (second row in the
Fig.~\ref{fig:lwi_noInitRef}). For the AMR simulation that started from the
coarsest level, we see that the diagonal jet is substantially shorter than that
of the reference solution and the AMR solution that has begun with the initial
refinement.
In order to prevent such coarse initial discretization error irreversibly
affecting an AMR simulation, small-scale features or discontinuities present in
the initial data should be resolved with a proper level of grid refinement if
possible.
A straightforward solution is to evaluate the AMR criteria on the initial data
and successively refine until reaching a consistently refined state, and then
begin the time evolution. For example, in the Liska--Wendroff implosion problem,
the initial grid refinement could be restricted to the discontinuous interface
along the diagonal \(x + y = 0.15\).

\section*{Acknowledgments}

We are grateful to Jacob Fields, James Stone, Romain Teyssier, and George Wong
for insightful discussions and suggestions. S.S. gratefully acknowledges support
from the Caltech Summer Undergraduate Research Fellowship (SURF) program. Y.K.
acknowledges support by the Sherman Fairchild Foundation, by the National
Science Foundation through the grants No. PHY-2309211, No. PHY-2309231, and No.
OAC-2209656 at Caltech, and by a joint fellowship at the Princeton Center for
Theoretical Science and Leinweber Forum for Theoretical Physics at Princeton
University. E.R.M. acknowledges support by the U.S. National Science Foundation
{(NSF)} under Grants No. AST-2508940 and  PHY-2541792. E.R.M. is also supported
by a Research Fellowship from the Sloan Foundation and a William H. Hurt
Scholarship at Caltech. The simulations presented in this article were performed
on computational resources managed and supported by Princeton University's
Research Computing and Northwestern Quest's High-Performance Computing Cluster.
Figures in this article were produced using \texttt{cmocean} \cite{cmocean},
\texttt{Matplotlib} \cite{the_matplotlib_development_team_2026_22700776}, and
\texttt{NumPy} \cite{numpy} packages. The authors acknowledge the usage of
Anthropic Claude for code debugging and assistance in formatting figures.

\bibliography{references}

\end{document}